\documentclass[journal,twocolumn,10pt]{IEEEtran}
\usepackage{graphicx}
\usepackage{amssymb}
\usepackage{mathtools}
\usepackage{dsfont}
\usepackage{cite}
\usepackage{stfloats}
\usepackage{subfigure}
\usepackage{psfrag}
\usepackage[mathscr]{euscript}
\usepackage{acronym}  
\usepackage{amsmath}
\usepackage{algorithm}
\usepackage[noend]{algpseudocode}
\usepackage{booktabs}
\usepackage{bbm}

\makeatletter
\def\BState{\State\hskip-\ALG@thistlm}
\makeatother

\usepackage{float}
\usepackage{balance}

\acrodef{CCDF}{complementary cumulative distribution function}
\acrodef{CF}{characteristic function}
\acrodef{PPP}{Poisson point processe}
\acrodef{RV}{random variable}
\acrodef{i.i.d.}{independent and identically distributed}
\acrodef{PDF}{probability distribution function}
\acrodef{CDF}{cumulative distribution function}
\acrodef{ch.f.}{characteristic function}
\acrodef{AWGN}{additive white Gaussian noise}
\acrodef{SNR}{signal-to-noise ratio}
\acrodef{LRT}{likelihood ratio test}
\acrodef{DRT}{distance ratio test}
\acrodef{GLRT}{generalized likelihood ratio test}
\acrodef{CRLB}{Cram\'{e}r-Rao lower bound}
\acrodef{CRB}{Cram\'{e}r-Rao bound}
\acrodef{ZZLB}{Ziv-Zakai lower bound}
\acrodef{ZZB}{Ziv-Zakai bound}
\acrodef{LOS}{line-of-sight}
\acrodef{ToF}{time-of-flight}
\acrodef{NLOS}{non-line-of-sight}
\acrodef{GDOP}{geometric dilution of precision}
\acrodef{GPS}{Global Positioning System}
\acrodef{FIM}{Fisher information matrix}
\acrodef{PEB}{position error bound}
\acrodef{SPEB}{squared position error bound}
\acrodef{TOA}{time-of-arrival}
\acrodef{TOF}{time-of-flight}
\acrodef{WSN}{wireless sensor network}
\acrodef{MAC}{medium access control}
\acrodef{RSS}{received signal strength}
\acrodef{WAF}{wall attenuation factor}
\acrodef{TDOA}{time difference-of-arrival}
\acrodef{RF}{radiofrequency}
\acrodef{RTT}{round-trip time}
\acrodef{AOA}{angle-of-arrival}
\acrodef{MF}{matched filter}
\acrodef{ED}{energy detector}
\acrodef{ML}{maximum likelihood}
\acrodef{MSE}{mean-square error}
\acrodef{RMSE}{root-mean-square error}
\acrodef{LEO}{localization error outage}
\acrodef{ppm}{part-per-million}
\acrodef{ACK}{acknowledge}
\acrodef{UWB}{Ultrawide bandwidth}
\acrodef{TNR}{threshold-to-noise ratio}
\acrodef{LS}{least squares}
\acrodef{IR-UWB}{impulse radio UWB}
\acrodef{FCC}{Federal Communications Commission}
\acrodef{TH}{time-hopping}
\acrodef{PPM}{pulse position modulation}
\acrodef{MUI}{multi-user interference}
\acrodef{PDP}{power delay profile}
\acrodef{BPZF}{band-pass zonal filter}
\acrodef{SIR}{signal-to-interference ratio}
\acrodef{SINR}{signal-to-interference-plus-noise ratio}
\acrodef{RFID}{radio frequency identification}
\acrodef{WPAN}{wireless personal area network}
\acrodef{WWB}{Weiss-Weinstein bound}
\acrodef{DP}{direct path}
\acrodef{MF}{matched filter}
\acrodef{MMSE}{minimum-mean-square-error}
\acrodef{SBS}{serial backward search}
\acrodef{SBSMC}{serial backward search for multiple clusters}
\acrodef{NBI}{narrowband interference}
\acrodef{WBI}{wideband interference}
\acrodef{INR}{interference-to-noise ratio}
\acrodef{CR}{channel response}
\acrodef{CIR}{channel impulse response}
\acrodef{CR}{channel  response}
\acrodef{RADAR}{radar}
\acrodef{MUR}{Multistatic radar}
\acrodef{JBSF}{jump back and search forward}
\acrodef{HDSA}{high-definition situation-aware}
\acrodef{RRC}{root raised cosine}
\acrodef{ST}{simple thresholding}
\acrodef{BTB}{Bellini-Tartara bound}
\acrodef{P-Max}{$P$-Max}  
\acrodef{MIMO}{multiple-input multiple-output}
\acrodef{MAP}{maximum a posteriori}
\acrodef{FG}{factor graph}
\acrodef{OP}{outage probability}
\acrodef{WED}{wall extra delay}
\acrodef{RMS}{root mean square}
\acrodef{SPAWN}{sum-product algorithm over a wireless network}
\acrodef{MDD}{minimum distance distribution}
\acrodef{MAP}{maximum a posteriori probability}
\acrodef{SAP}{small cell access point}
\acrodef{UE}{user equipment}
\acrodef{MBS}{macro cell base station}
\acrodef{UER}{\ac{UE} Relay}
\acrodef{D2D}{device-to-device}
\acrodef{MBS}{macro base station}
\acrodef{CSI}{channel state information}
\acrodef{OGR}{outage guard region}
\acrodef{FUR}{feasible UER region}
\acrodef{EHR}{energy harvesting region}
\acrodef{EH}{energy harvesting}
\acrodef{D2D-EHSN}{D2D communication provided \ac{EH} small cell network}
\acrodef{D2D-EHHN}{D2D communication provided \ac{EH} heterogeneous network}
\acrodef{3GPP}{3rd Generation Partnership Project}
\acrodef{BS}{base station}
\acrodef{DF}{decode and forward}
\acrodef{CCDF}{complementary cumulative distribution function}
\acrodef{ZF}{zero forcing}
\acrodef{RZF}{regularized zero forcing}
\acrodef{WLLN}{weak law of large number}
\acrodef{SLLN}{strong law of large numbers}
\acrodef{TDD}{Time-division duplex}
\acrodef{EE}{energy efficiency} 
\acrodef{HetNet}{heterogeneous network} 
\acrodef{SCP}{Single Cell Processing}
\acrodef{CBF}{Coordinated Beamforming}
\usepackage{color}
\usepackage{dsfont}
\usepackage{bbm}

\def\PST{P_{\mathrm{st}}}

\DeclareMathAlphabet{\mathsf}{OML}{cmbr}{m}{it}

\newtheorem{theorem}{\bf Theorem}
\newtheorem{lemma}{\bf Lemma}

\newtheorem{remark}{\bf Remark}

\newcommand{\bd}{\begin{description}}
\newcommand{\ed}{\end{description}}
\newcommand{\be}{\begin{enumerate}}
\newcommand{\ee}{\end{enumerate}}
\newcommand{\bi}{\begin{itemize}}
\newcommand{\ei}{\end{itemize}}
\newcommand{\bl}{\begin{list}}
\newcommand{\el}{\end{list}}
\newcommand{\bt}{\begin{tabbing}}
\newcommand{\et}{\end{tabbing}}

\newcommand{\paperTitle}{ Locally Adaptive Scheduling Policy for Optimizing Information Freshness in Wireless Networks }

\begin{document}

{
\title{\paperTitle}

\author{

	    Howard~H.~Yang$^\ast$, Ahmed~Arafa$^\dagger$,
	    Tony~Q.~S.~Quek$^\ast$, and H. Vincent Poor$^\dagger$ \\
       $^\ast$\textit{Singapore University of Technology and Design, Singapore}\\
       $^\dagger$\textit{Princeton University, NJ, USA}

 \thanks{  This work was supported in part by the U.S. National Science Foundation under Grants CCF-0939370 and CCF-1513915. }

%
}
\maketitle
\acresetall
\thispagestyle{empty}
\begin{abstract}
Optimization of information freshness in wireless networks has usually been performed based on queueing analysis that captures only the temporal traffic dynamics associated with the transmitters and receivers. However, the effect of interference, which is mainly dominated by the interferers' geographic locations, is not well understood.
In this paper, we leverage a spatiotemporal model, which allows one to characterize the age of information (AoI) from a joint queueing-geometry perspective, and design a decentralized scheduling policy that exploits local observation to make transmission decisions that minimize the AoI.
Simulations results reveal that the proposed scheme not only largely reduces the peak AoI but also scales well with the network size.
\end{abstract}

\acresetall
\section{Introduction}\label{sec:intro}
Fast growing wireless services like factory automation and vehicular communication, as well as the likes of mobile applications, have imposed more stringent requirement for the timely delivery of information.
To give an adequate response, network operators need not only understand how the network activities affect the timeliness of information delivery, but more importantly, they need to assert substantial control to enhance transmission.
Recognizing the limitation in conventional performance indicators, e.g., delay or throughput, as not being able to account the ``information lag'' caused by queueing aspects, there emerges a new metric, referred to as the \textit{age of information (AoI)}, which explicitly measures the time elapsed since the last recorded  update was generated \cite{KauYatGru:12}.
From then on, a series of research has been conducted to seek different approaches, mainly in the form of scheduling protocols, to optimize information freshness in the context of wireless networks \cite{HeYuaEph:16,KadUysSin:16,TalKarMod:18,talak2018optimizing}.
The problem of finding optimal scheduling protocol, despite being NP hard \cite{HeYuaEph:16}, is shown to possess a solution in terms of a greedy algorithm, which  schedules the link with highest age to transmit, in a symmetric network \cite{KadUysSin:16}.
Moreover, depending on whether the channel state is perfectly available \cite{TalKarMod:18} or not \cite{talak2018optimizing}, advanced virtual queue and age based protocols are proposed.
However, these models simplify the packet departure process by adopting a Poisson process, and do not account for the interference that differs according to distance between simultaneous transmitters as well as channel gains.
As a result, the information-theoretic interactions are yet to be precisely captured.

By nature, the wireless channel is a broadcast medium. Thus, transmitters share a common spectrum in space will interact with each other through the interference they cause.
To understand the performance of communication links in such networks, stochastic geometry has been introduced as a tool by which one can model node locations as spatial point processes and obtain closed form expressions for various network statistics, e.g., the distribution of interference, the successful transmission probability, and the coverage probability \cite{BacBla:09}.
The power of stochastic geometry has made it a disruptive tool for performance evaluation among various wireless applications, including ad-hoc and cellular networks \cite{AndBacGan:11}, D2D communications \cite{YanLeeQue:16}, MIMO \cite{YanGerQue:16}, and mmWave systems \cite{BaiHea:15}.
While such model has been conventionally relying on the \textit{full buffer} assumption, i.e., every link always has a packet to transmit, a line of recent works managed to bring in queueing theory and relax this constraint \cite{GhaElsBad:17,ZhoQueGe:16,YanQue:19,YanWanQue:18}, allowing one to give a complete treatment for the behavior of wireless links from both spatial and temporal perspectives. As a result, the model is further employed to design scheduling policies \cite{ZhoQueGe:16,YanWanQue:18}, study the scaling property in IoT networks \cite{GhaElsBad:17}, and analyze the delay performance in cellular network \cite{YanQue:19}.
In this paper, we leverage a spatiotemporal model as in \cite{ChiElSCon:17} for the design of a transmission protocol that optimizes information freshness in wireless networks.
Particularly, we model the deployment of transmitters and receivers as independent Poisson point processes (PPPs).
The temporal dynamic of AoI is modelled as a discrete-time queueing system, in which we consider the arrival of packets at each transmitter to be independent Bernoulli processes.
Each transmitter maintains an infinite capacity buffer to store the incoming packets, and initiates a transmission attempt at each time slot with some probability if the buffer is not empty.
Transmissions are successful only if the signal-to-interference-plus-noise ratio (SINR) exceeds a predefined threshold, upon which the packet can be removed from the buffer.
We propose a decentralized scheduling policy to minimize the AoI in a wireless network. The proposed scheme is efficient in the sense that it requires only local information and has very low implementation complexity.
Simulation results demonstrate the effectiveness of our proposed scheme in reducing the peak AoI. Moreover, the proposed scheme is also shown to adequately adjust according to the change of the ambient environment and thus scales well as the network grows in size.

\section{System Model}\label{sec:sysmod}

We model the wireless network as a set of transmitter-receiver pairs, all located in the Euclidean plane. The transmitting nodes are scattered according to a homogeneous Poisson point process (PPP) $\tilde{\Phi}$ of spatial density $\lambda$. Each transmitter $X_i \in \tilde{\Phi}$ has a dedicated receiver, whose location $y_i$ is at distance $r$ in a random orientation. According to the displacement theorem \cite{BacBla:09}, the location set $\bar{\Phi} = \{y_i\}_{i=0}^\infty$ also forms a homogeneous PPP with spatial density $\lambda$.
We segment the time into slots with the duration of each slot equal to the time to transmit a single packet.
The packet arrival process at each transmitter is modeled as independent and identically distributed (i.i.d.) Bernoulli with parameter $\xi$.
All incoming packets are stored in a single-server queue with infinite capacity under the first-come-first-serve (FCFS) discipline.
During each time slot, the queue-nonempty transmitter will initiate a channel access attempt according to its scheduling protocol, and send out one packet upon approval. The transmission succeeds if the signal-to-interference-plus-noise ratio (SINR) at the corresponding receiver exceeds a predefined threshold.
A packet is removed from the buffer when its reception is acknowledged by the receiver through an ACK feedback.
If the packet is not correctly decoded, the receiver sends a NACK message and the packet is retransmitted. We assume the ACK/NACK transmission is instantaneous and error-free, as commonly done in the literature \cite{TalKarMod:18}.
In order to investigate the time domain evolution, we limit the mobility of transceivers by considering a static network, i.e., the locations of transmitters and receivers remain unchanged in all the time slots.

We assume that each transmitter uses unit transmission power $P_{\mathrm{tx}}$. The channel is subjected to both Rayleigh fading, which varies independently across time slot, and path-loss that follows power law attenuation.
Moreover, the receiver is also subjected to white Gaussian thermal noise with variance $\sigma^2$. By applying Slivnyak's theorem \cite{BacBla:09}, it is sufficient to focus on a \textit{typical} receiver $y_0$ located at the origin, with its tagged transmitter at $X_0$. Thus, when the tagged transmitter sends out a packet during slot $t$, the corresponding SINR received at the typical node can be written as
\begin{align}
\mathrm{SINR}_{0,t} = \frac{P_{\mathrm{tx}} H_{00} r^{-\alpha} }{ \sum_{ j \neq 0 } P_{\mathrm{tx}} H_{j0} \zeta_{j,t} \nu_{j,t} \Vert X_j - y_0 \Vert^{-\alpha} + \sigma^2 }
\end{align}
where $\alpha$ denotes the path loss exponent, $H_{ji} \sim \exp(1)$ is the channel fading from transmitter $j$ to receiver $i$, $\zeta_{j,t} \in \{ 0, 1 \}$ is an indicator showing whether the buffer of node $j$ is empty ($\zeta_{j,t}=0$) or not ($\zeta_{j,t}=1$), and $\nu_{j,t} \in \{ 0, 1 \}$ represents the scheduling decision of node $j$, where it is set to 1 upon assuming transmission approval and 0 otherwise.

\subsection{Age of Information}
Without loss of generality, we denote the communication link between the transmitter-receiver pair located at $(X_0, y_0)$ as \textit{typical}.
Then, as illustrated in Figure~\ref{fig:AoIMod_V1}, the AoI $A_0(t)$ over the typical link grows linearly in the absence of successful communication, and, when the transmission is successful, reduces to the time elapsed since the generation of the delivered packet. To make the statement more precise, we formalize the evolution of $A_0(t)$ via the following expression
\begin{align*}
A_0(t \!+\! 1) \! = \!
\left\{
       \begin{array}{ll}
         \!\!\! A_0(t) \!+\! 1, ~\quad\quad  \text{if transmission fails}, \\
         \!\!\! t \!-\! G_0(t) \!+\! 1, \quad \text{otherwise}
       \end{array}
\right.
\end{align*}
where $G_0(t)$ is the generation time of the packet delivered over the typical link at time $t$.

In the sequel, we use the \textit{peak AoI} as our metric to evaluate the age performance across a wireless network. Formally, the peak AoI at one generic link $j$ is defined as
\begin{align}
A^{ \mathrm{p} }_j = \limsup\limits_{ N \rightarrow \infty } \frac{ \sum_{n=1}^N A_j( T_j(n) ) }{N},
\end{align}
where $T_j(n)$ is the time slot at which the $n$-th packet from link $j$ is successfully delivered. We can extend this concept to a network scale and define the \textit{network} peak AoI as follows
\begin{align*}
A^{\mathrm{p}} &= \limsup_{R \rightarrow \infty} \frac{ \sum_{ X_j \in \tilde{\Phi} \cap B(0,R) } A^{ \mathrm{p} }_j }{ \sum_{ X_j \in \tilde{\Phi} } \chi_{ \{ X_j \in B(0,R) \} }  }\\
&\stackrel{(a)}{=}   \mathbb{E}^0 \Big[ \limsup_{ N \rightarrow \infty } \frac{1}{N} \sum_{n=1}^N A_0( T_0(n) )  \Big]
\end{align*}
where $B(0,R)$ denotes a disk centered at the origin with radius $R$, $\chi_{E}$ is an indicator function which takes value 1 if event $E$ occurs and 0 otherwise, and $(a)$ follows from the Campbell's theorem \cite{BacBla:09}.
The notion $\mathbb{E}^0[\cdot]$ indicates the expectation is taken with respect to the Palm distribution $\mathbb{P}^0$ of the stationary point process, where under $\mathbb{P}^0$ almost surely there is a node located at the origin \cite{BacBla:09}.

\begin{figure}[t!]
  \centering{}

    {\includegraphics[width=0.95\columnwidth]{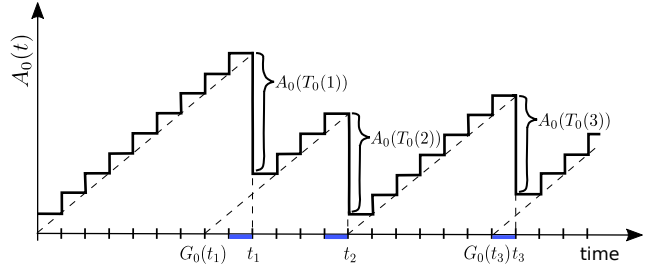}}

  \caption{ An example of the time evolution of age at a typical link. The time instances $G_0(t_i)$ and $t_i$ respectively denote the moment when the $i$-th packet is generated and delivered, and the age is reset to $t_i - G_0(t_i) + 1$. Here, $t_i = T_0(i)$ with $T_0(i)$ defined in (2). }
  \label{fig:AoIMod_V1}
\end{figure}

\subsection{Stopping Sets and Scheduling Policy}
In a wireless network, as all transmitters are intertwined through the interference they cause to each other, it is important to have an effective protocol that schedules the appropriate channel access state for each node.
Inspired by the fact that knowledge from local activities can be utilized to improve the overall network performance, we incorporate such local information in the design of the scheduling policy.

Since a generic transmitter usually has limited sensing power whereas it can only obtain the information about its geometry vicinity, we encapsulate such local knowledge by the notion of \textit{stopping set} $S = S(\tilde{\Phi}, \bar{\Phi})$ \cite{BacBla:09,BacBlaSin:14}.
More precisely, the stopping set is a random element taking each realization from the Borel sets in $\mathbb{R}^2$ such that for any observation window $A$, one can determine whether $S = S(\tilde{\Phi}, \bar{\Phi}) \subset A$.
This concept enables us to model the region in which the information of nodes, including their locations, are known to a typical observer.
In particular, depending on the scenarios under consideration, stopping sets can take various forms.
For instance, if the transmitters have unified sensing power, the observation region at each node will be a
disk with constant radius and the stopping set takes a deterministic
form. When the transmitters are empowered with heterogeneous
sensing capabilities, each node may want to obtain information up to
the $p$-th nearest neighbor, in which case the observation region varies across different nodes and the stopping set takes a random shape. Aside from disks, the stopping set can have more complicated formats, e.g.,
a hexagon under clustering regulation or different orders of Voronoi cells in the context of cellular
networks \cite{YanGerQue:17}, depending on the specific task under consideration.
An illustration of deterministic stopping sets in a Poisson bipolar network is given in Fig.~\ref{fig:StpSet_V1}. Note that different transmitters, e.g., the ones located at $X_1$ and $X_2$, can have various local observations.

To generalize the concept network wide, we further introduce a shifting operation, denoted by $\theta_x$ and performs on $\tilde{\Phi}$ and $\bar{\Phi}$, which translates all the network nodes by the vector $-x$, i.e., $\theta_{x}\{X_i\} = \{ X_i - x \}$ and similarly for the receivers.
Extending this operator to all subsets $A \subset \mathbb{R}^2$, we have $\theta_x(A) = \{ a - x: a \in A \}$.
To this end, we are able to construct a \textit{translation invariant} policy where all nodes $i$ set their activation probability to $\gamma_i = \eta_{\mathrm{S}}(\theta_{X_i}\!\tilde{\Phi}, \theta_{X_i}\!\bar{\Phi} )$, with $\eta_{\mathrm{S}}(\cdot)$ being a measurable function whose argument is the network geometry ($\tilde{\Phi}$, $\bar{\Phi}$) and has value in $[0, 1]$.
In other words, any node $i$, in order to choose its channel access probability, applies the policy $\eta_{\mathrm{S}}$ evaluated for the knowledge learned from its geometric proximities in $S$.
For a given stopping set $S = S(\tilde{\Phi}, \bar{\Phi})$, we consider the following class of scheduling policies with local spatial information $S$:
\begin{align}
\eta_{\mathrm{S}}(\tilde{\Phi}, \bar{\Phi}) = \eta_{\mathrm{S}}\!\left( \tilde{\Phi} \cap S, \bar{\Phi} \cap S \right).
\end{align}
Note that to apply such a policy and evaluate the scheduling policy $\eta_{\mathrm{S}}( \theta_{X_i}\! \tilde{\Phi}, \theta_{X_i}\! \bar{\Phi} )$, node $i$ needs to obtain only local knowledge about the other nodes in the stopping set $S_i = {S}(\theta_{X_i}\! \tilde{\Phi}, \theta_{X_i}\! \bar{\Phi})$.
In this regard, the scheme can run without a central controller and is thus decentralized.

\remark{
\textit{
	By leveraging the notion of stopping sets, our framework is able to provide a unified approach to account for various types of local information.
}
 }

\begin{figure}[t!]
  \centering{}

    {\includegraphics[width=0.95\columnwidth]{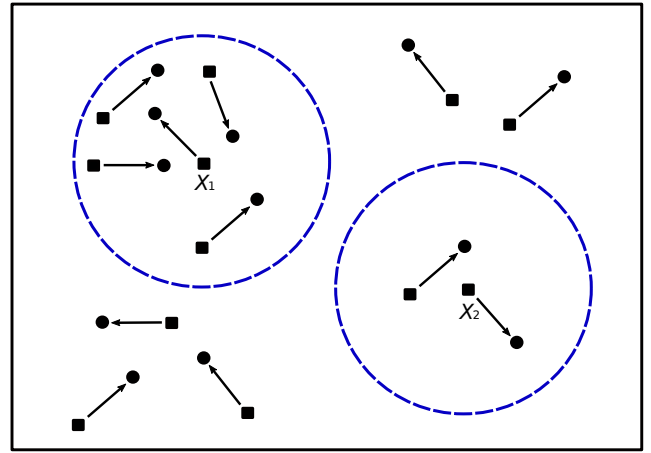}}

  \caption{ Illustration of a Poisson bipolar network with stopping sets being disks with constant radii, where black squares and dots are the transmitters and receivers, respectively, and the circles with dashed lines are two exemplary stopping sets centered at $X_1$ and $X_2$. }
  \label{fig:StpSet_V1}
\end{figure}

\section{Scheduling Policy Design}
\subsection{Preliminaries}
The radio interface between any transmitter-receiver pair $i$ can be modeled as a Geo/G/1 queue where the departure rate varies according to the link throughput.
In the steady state, the link throughput, or equivalently service rate, is determined by both the scheduled channel access probability, i.e., how frequent a transmitter with non-empty buffer can access the channel, and the transmission success probability. Particularly, conditioned on the realization of the point process  $\Phi \triangleq \tilde{\Phi} \cup \bar{\Phi}$, the  transmission success probability, $\mu^\Phi_i$, is given by \cite{ZhoQueGe:16,YanQue:19}\footnote{ In the following, we will drop the time index $t$ from the subscript as we are dealing with the situation under steady state. }
\begin{align} \label{equ:TXSucProb}
\mu_i^\Phi = \mathbb{P}\left( \mathrm{SINR}_i > T | \Phi  \right)
\end{align}
where $T$ is the decoding threshold.
The following lemma characterizes the impact of the transmission status from the interfering nodes on a typical link:
\begin{lemma}
\textit{
	Conditioned on the spatial realization $\Phi$, the transmission success probability at the typical link is given by
\small
	\begin{align} \label{equ:CondThrPut}
	\mu_0^\Phi = e^{-\frac{ T r^\alpha }{\rho}}  \prod_{j \neq 0} \bigg( 1 - \frac{ a_j^\Phi \gamma_j^\Phi }{ 1 + \mathcal{D}_{j0} } \bigg)
	\end{align}
\normalsize
	where $\rho = P_\mathrm{tx}/\sigma^2$, $a_j^\Phi = \lim_{ t \rightarrow \infty } \mathbb{P}( \zeta_{j,t} = 1 | \Phi )$, $\gamma_j^\Phi = \lim_{ t \rightarrow \infty } \mathbb{P}(\nu_{j,t} = 1 | \zeta_{j,t} = 1, \Phi )$, and $\mathcal{D}_{ij} = \Vert X_i - y_j \Vert^\alpha / T r^\alpha$.
}
\end{lemma}
\begin{IEEEproof}
Conditioned on the spatial realization of all the transceiver locations, the transmission success probability can be derived as follows:
\begin{align*}
&\mathbb{P}\!\left( \mathrm{SINR}_0 \!>\! T | \Phi  \right)
\!=\! \mathbb{P} \bigg(  \frac{ H_{00} }{ T r^\alpha }  >  \sum_{ j \neq 0 } \frac{ H_{j0} \zeta_{j,t} \nu_{j,t} }{ \Vert X_j \!-\! y_0 \Vert^\alpha } \!+\! \frac{1}{\rho} \,  \Big| \, \Phi  \bigg)
\nonumber\\
&= \mathbb{E}\bigg[ e^{-\frac{ T r^\alpha }{\rho}} \prod_{ j \neq 0 } \exp\! \Big(\! - T r^\alpha \frac{ H_{j0} \zeta_{j,t} \nu_{j,t} }{ \Vert X_j \!-\! y_0 \Vert^\alpha } \Big)  \Big| \Phi \bigg]
\nonumber\\
&\stackrel{(a)}{=} e^{-\frac{ T r^\alpha }{\rho}}  \prod_{j \neq 0} \Big( 1 - a_j^\Phi \gamma_j^\Phi + \frac{ a_j^\Phi \gamma_j^\Phi }{ 1 + 1/\mathcal{D}_{j0} } \Big),
\end{align*}
where $(a)$ is by noticing that $\mathbb{P}( \zeta_{j,t} \nu_{j,t} = 1 | \Phi ) = \mathbb{P}( \zeta_{j,t} = 1 | \Phi ) \times \mathbb{P}( \nu_{j,t} = 1 | \zeta_{j,t} = 1, \Phi ) $, and the result follows from further simplifying the product factors.
\end{IEEEproof}

Next, by conditioning on the realization of the point process $\Phi$, the communication between a typical transceiver pair can be regarded as a Geo/Geo/1 queue where the service rate is given by $\gamma_0^\Phi \mu_0^\Phi$.
As such, using tools from queueing theory, we arrive at a conditional form of the peak AoI.
\begin{lemma}\label{lma:Cond_AoI}
\textit{
	In the steady state, conditioned on the spatial realization $\Phi$, the peak AoI at a typical link is given as
	\begin{align} \label{equ:CondPeakAoI}
  \mathbb{E}^0\!\left[ A^{\mathrm{p}} | \Phi \right] = \left \{ \!\!\!
    \begin{tabular}{cc}
    $ \frac{1}{\xi} +\!  \frac{ 1 - \xi }{ \gamma_0^\Phi \mu_0^\Phi - \xi }$, & if $\, \gamma_0^\Phi \mu_0^\Phi  > \xi$,   \\
    +$\infty$, &  if $\, \gamma_0^\Phi \mu_0^\Phi \leq \xi$.
    \end{tabular}
    \right.
	\end{align}
}
\end{lemma}
\begin{IEEEproof}
Let us denote by $M_t$ and $N_t$ the inter-arrival time and the total sojourn time in the queue, respectively. The conditional peak AoI is then given by \cite{HuaMod:15}
\begin{align} \label{equ:Cndtn_Peak}
\mathbb{E}^0\! \left[ A^{\mathrm{p}} | \Phi \right] = \mathbb{E}^0\!\left[ M_t + N_t | \Phi \right] = \mathbb{E}\!\left[ M_t \right] + \mathbb{E}^0\!\left[ N_t | \Phi \right].
\end{align}
On the one hand, with packet arrivals following the Bernoulli distribution, which is independent with the departure process, we have $\mathbb{E}[M_t] = 1/\xi $.
On the other hand, the average sojourn time of a Geo/Geo/1 queue can be calculated as \cite{YanWanQue:18}
\begin{align*}
\mathbb{E}^0\!\left[ N_t | \Phi \right] = \left \{ \!\!\!
\begin{tabular}{cc}
$\frac{ 1 - \xi }{ \gamma_0^\Phi \mu_0^\Phi - \xi }$, & if $~ \gamma_0^\Phi \mu_0^\Phi > \xi$,   \\
$+ \infty$, &  if $~ \gamma_0^\Phi \mu_0^\Phi \leq \xi$
\end{tabular}
\right.
\end{align*}
where $\gamma_0^\Phi \mu_0^\Phi > \xi$ is the break event for queueing stability. The result then follows from substituting the above results back into \eqref{equ:Cndtn_Peak}.
\end{IEEEproof}

From \eqref{equ:CondPeakAoI}, it is obvious that the principle of optimizing information freshness consists in maximizing the link throughput.
In order to achieve this goal, a policy that schedules the channel access at each node by jointly balancing the radio resource utility and mutual interference is essential.
In the following, we formulate a stochastic decision problem to find the scheduling policy that accomplishes this task.
\subsection{Locally Adaptive Scheduling Policy}
\subsubsection{Design} Let a stopping set $S = S(\tilde{\Phi}, \bar{\Phi})$ be given. Using Lemma~\ref{lma:Cond_AoI}, the design of scheduling policy can be written as:
\begin{align} \label{prbm:PeakAoI}
& \min_{ \eta_{\mathrm{S}} } ~~~~ \mathbb{E}^0_\Phi\! \left[ \frac{ 1 - \xi }{ \gamma_0^\Phi \mu_0^\Phi - \xi } \right] + \frac{1}{\xi}\\ \label{equ:Rst_SpSt}
& ~~ \mathrm{s.t.} \quad~ 0 \leq \gamma_i^\Phi = \eta_{\mathrm{S}}\!\left( \theta_{X_i} \! \tilde{\Phi}, \theta_{X_i}\! \bar{\Phi} \right)  \leq 1, \\ \label{equ:Rst_Stb}
& \qquad \quad ~ \xi \leq \mathbb{E}^0_{\Phi}[ \eta_{\mathrm{S}} \mu_0^\Phi ].
\end{align}
It is worthwhile to point out that the design factor $\eta_{ \mathrm{S} }$ in \eqref{prbm:PeakAoI} is not a single parameter but instead a policy, which takes input the state information, i.e., the node's location and information observed from the corresponding stopping set, and as an output the channel access probability.
As such, the scheduling policy varies from node to node, which is stated in constraint \eqref{equ:Rst_SpSt}, because the local knowledge is location dependent.
Moreover, the queueing stability shall be guaranteed in the average sense, as shown in \eqref{equ:Rst_Stb}, according to the Loynes' theorem \cite{YanQue:19}.

By taking the expression of $\mu_0^\Phi$ as in \eqref{equ:CondThrPut}, we note that the optimization of \eqref{prbm:PeakAoI} requires the typical link to have information regarding: $i$) parameters $\mathcal{D}_{0j}$, which depend on the location of receivers in $\bar{\Phi} \cap S $, and $ii$) their corresponding active states.
However, due to the interaction of queueing dynamics at different nodes, the active state at each transmitter varies over time and is difficult to keep track with. As such, we leverage the conventional dominant system argument \cite{ZhoQueGe:16}, where each transmitter keeps sending out packets in each time slot (if one transmitter has empty buffer at some time slot, it sends a dummy packet), and minimizes the following alternative:
\begin{align} \label{prbm:Dom_PeakAoI}
& \min_{ \eta_{\mathrm{S}} } ~~~~ \mathbb{E}^0_\Phi\! \left[ \frac{ 1 - \xi }{ \gamma_0^\Phi \hat{\mu}_0^\Phi - \xi } \right] + \frac{1}{\xi}\\ \label{equ:Dom_Rst_SpSt}
& ~~ \mathrm{s.t.} \quad~ 0 \leq \gamma_i^\Phi = \eta_{\mathrm{S}}\!\left( \theta_{X_i} \! \tilde{\Phi}, \theta_{X_i}\! \bar{\Phi} \right),
\end{align}
where $\hat{\mu}_0^\Phi$ is given by taking $a_j^\Phi = 1$ in \eqref{equ:CondThrPut}, i.e.,
\begin{align} \label{equ:Dmt_CondThrPut}
\hat{\mu}_0^\Phi = e^{-\frac{ T r^\alpha }{\rho}}  \prod_{j \neq 0} \bigg( 1 - \frac{ \gamma_j^\Phi }{ 1 + \mathcal{D}_{j0} } \bigg).
\end{align}
In this regard, the design of the scheduling policy is free from tracking the active state of different transmitting nodes.
That brings us to the main structural result of this paper.
\begin{theorem}\label{thm:DmSym_PeakAoI}
\textit{
    For all given stopping sets $S = S(\tilde{\Phi}, \bar{\Phi})$, the solution to the optimization problem in \eqref{prbm:Dom_PeakAoI} is given by the solution of the following fixed point equation
	\begin{align}\label{equ:OptSln}
	\frac{1}{\eta_{\mathrm{S}}} -\!\!\!\!\! \sum_{ \substack{ j \neq 0, y_j \in S } } \frac{1}{ 1  \!+\! \mathcal{D}_{0j} \!-\! \eta_{\mathrm{S}} } -\!\! \int_{\mathbb{R}^2 \setminus S }\! \frac{ \lambda  dz }{ 1 \!+\! \Vert z \Vert^\alpha \! / T r^\alpha } = 0
	\end{align}
	if the following holds
	\begin{align} \label{equ:CndOptl}
	\sum_{ \substack{ j \neq 0, y_j \in S } }  \frac{1}{ \mathcal{D}_{0j} } +\! \int_{ \mathbb{R}^2 \setminus S }\! \frac{ \lambda dz }{ 1 \!+\! \Vert z \Vert^\alpha \! / T r^\alpha  } > 1.
	\end{align}
	Otherwise, $\eta_{\mathrm{S}} = 1$.
}
\end{theorem}
\begin{IEEEproof}
See Appendix~\ref{apx:OptCtrl_PeakAoI}.
\end{IEEEproof}

The result from \eqref{equ:OptSln} gives an explicit form to calculate the scheduling policy at the typical node $0$, i.e., $\gamma_0 = \eta_{ \mathrm{S} }(\theta_{ X_0 } \tilde{\Phi}, \theta_{ X_0 } \bar{\Phi} )$.
In terms of a generic node $i$, the scheduling policy can be attained by shifting the origin of the point process $\Phi$ to $X_i$ and then apply the above result, i.e., $\gamma_i = \eta_{ \mathrm{S} }(\theta_{ X_i } \tilde{\Phi}, \theta_{ X_i } \bar{\Phi} )$.
Moreover, note from \eqref{equ:CndOptl} that the transmitters will take an opportunistic channel access approach when the following holds:
\begin{align*}
T &> \frac{ { r^{ - \alpha} } }{ \sum\limits_{ j \neq 0, y_j \in S } \!\!\!\!\! { \Vert y_j \Vert^{-\alpha} } \!+\! \int_{ \mathbb{R}^2 \setminus S } \frac{ \lambda dz }{ \Vert z \Vert^\alpha  } } \\
& = \frac{ \mathbb{E} \big[ P_{\mathrm{tx}} H_{00} r^{-\alpha} \big] }{ \mathbb{E}\big[ \sum_{ j \neq 0 } P_{\mathrm{tx}} H_{0j} \Vert X_0 - y_j \Vert^{-\alpha} | S \big] },
\end{align*}
which implies that a transmitter should reduce the frequency of channel access when the average ratio between its signal power and the interference it generates into the network is smaller than the SINR threshold.
Such approach coincides with the intuition that transmitters located close to each other can cause severe mutual interference and need to be scheduled for the channel access, while the ones located far away from their neighbors can access the radio channel more frequently.

\subsubsection{Examples} Below, we illustrate the proposed scheme via a few examples to better understand the results of the theorem. To keep the results intuitive, we limit the example stopping sets to be disk-based, but note that the framework is quite versatile and can accommodate more general situations, e.g., with stopping sets being the irregular extended cells \cite{YanGerQue:17}.

$a) ~S = \emptyset$: When transmitters have no topological information about their neighbors, the scheduling policy shall be assigned as a universal constant. Using results from Theorem~\ref{thm:DmSym_PeakAoI}, we have $\eta_{ \mathrm{S} }(X_i)=1, \forall i \in \mathbb{N}$.
This can also be recognized by the fact that the average link throughput
\begin{align*}
\mathbb{E}^0[ \gamma_0 \hat{\mu}_0 ] = \eta_{ \mathrm{S} } \exp\big( - \int_0^\infty \!\!\! \frac{ 2 \pi \lambda \eta_{ \mathrm{S} } v dv }{ 1 + v^\alpha / Tr^\alpha } \big)
\end{align*}
monotonically increases with respect to $\eta_{ \mathrm{S} }$. Thus, the best strategy is to set $\eta_{\mathrm{S}} = 1$ at every transmitter.

$b)~ S = B(0,\Vert y_\mathrm{c} \Vert)$: Here, for a generic transmitter $j$, $y_{\mathrm{c}}$ denotes the nearest receiver that node $j$ generates interference to.
This corresponds to the scenario where $S$ is a random stopping set.
By solving \eqref{equ:OptSln}, the scheduling policy takes the following form:
\begin{align*}
{\eta_{\mathrm{S}}} \!=\! \frac{ 1 }{ \mathcal{M}( \Vert y_\mathrm{c} \Vert ) } \!+\! \frac{ \Vert y_\mathrm{c} \Vert^\alpha \!\!+\! T r^\alpha }{ 2 T r^\alpha } \!-\! \sqrt{ \frac{ T r^\alpha \!+\! \Vert y_\mathrm{c} \Vert^\alpha  }{ 4 T r^\alpha } \!+\! \frac{ 1 }{ \mathcal{M}( \Vert y_\mathrm{c} \Vert )^2 } },
\end{align*}
where $\mathcal{M}( \Vert y_\mathrm{c} \Vert )$ is given as
\begin{align*}
\mathcal{M}( \Vert y_\mathrm{c} \Vert ) = \int_{ \Vert y_\mathrm{c} \Vert }^\infty \frac{ 2 \pi \lambda v dv }{ 1 + v^\alpha / T r^\alpha }.
\end{align*}

$c)~ S = B(0,R)$: In this example, each transmitter knows the location of receivers in a disk of radius $R$ centered at its location, with $R>0$. Note that such stopping set is a deterministic set. The proposed policy can thus be attained by solving
\begin{align*}
\frac{1}{ \eta_{\mathrm{S}} } =\!\!\!\!\! \sum_{ 0 < \Vert y_j \Vert \leq R } \frac{1}{ \frac{ \Vert y_j \Vert^\alpha }{ T r^\alpha } \!+\! 1 \!-\! \eta_{\mathrm{S}}  } +  2 \pi \! \int_{R}^\infty \! \frac{  \lambda v dv }{ 1 \!+\! { v^\alpha } / { T r^\alpha } }.
\end{align*}


\section{Simulation and Numerical Results}
We evaluate the performance of the proposed scheduling policy in this section.
Particularly, during each simulation run, we realize the locations of transmitters and receivers over a 10 $\text{km}^2$ area via homogeneous PPPs. The packets arriving at each node are generated according to independent Bernoulli processes. We average over 10,000 realizations and collect the statistic from each communication link to finally obtain the peak AoI.
Unless differently specified, we use the following parameters: $\alpha = 3.8$, $T=0$~dB, $\PST=23.7$~dBm, $\sigma^2 = -90$~dBm, and $\lambda = 10^{-4}\mathrm{m}^{-2}$.

\begin{figure}[t!]
  \centering{}

   \subfigure[\label{fig:1a_prim}]{\includegraphics[width=0.95\columnwidth]{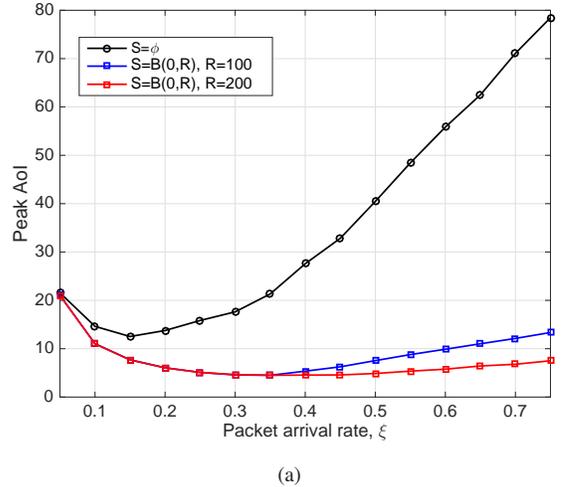}}
   \subfigure[\label{fig:1b_prim}]{\includegraphics[width=0.95\columnwidth]{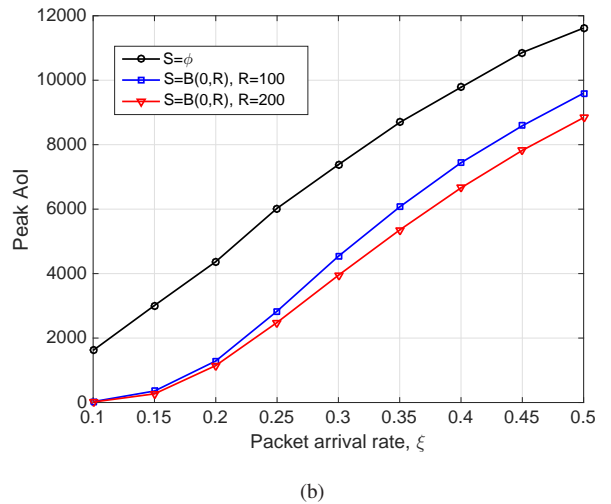}}

  \caption{ Peak AoI vs packet arrival rate: (a) $r=25$, with deterministic stopping set $S = B(0,R)$, (b) $r=100$, with deterministic stopping set $S = B(0,R)$. }
  \label{fig:PeakAoI_vs_xi}
\end{figure}

Fig.~\ref{fig:PeakAoI_vs_xi} compares the proposed scheduling policy with local observation from a deterministic stopping set, i.e., $S=B(0,R)$ where $R$ is a constant, to that with no available local information, i.e., $S=\phi$ (in which case, $\eta_{\mathrm{S}} =1$, $\forall j \in \mathbb{N}$), under different transmitter-receiver distances.
From Fig.~\ref{fig:1a_prim}, we immediately note an optimal packet arrival rate exists for both cases due to a tradeoff between update frequency and the incurred delay.
Moreover, the figure also shows that once armed with sufficient local information, the proposed scheduling method is able to maintain the AoI at low level for a wide range of packet arrival rates, demonstrating its effectiveness in optimizing the information freshness in wireless networks.
On the other hand, from Fig.~\ref{fig:1b_prim} we observe that in regimes with weak signal power (i.e., the distance between each transmitter-receiver pair is large), the proposed scheduling policy greatly improves the peak AoI compared to that without local information.
The gain is especially remarkable in the regime with low to moderate packet arrival rates, in which the proposed scheme reduces more than a half the peak AoI. This is mainly because such regime is where the SINR rapidly degrades while only few packets are accumulated in the buffer, and hence if transmitters can tolerate certain increments in the queueing delay and control their channel access frequency, the SINR can be greatly boosted up and it results in a much shorter transmission delay.
We further note that the performance of the proposed scheme can be enhanced through expanding the observation region, i.e., by increasing the radius of the deterministic disk. Therefore, the amount of local observation plays a vital role in the scheduling design.

\begin{figure}[t!]
  \centering{}

    {\includegraphics[width=0.95\columnwidth]{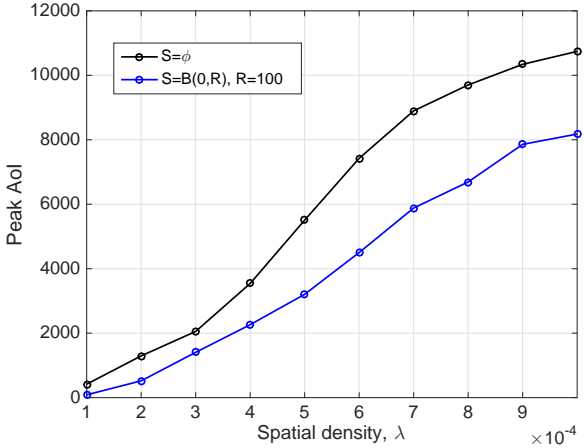}}

  \caption{ Peak AoI vs spatial density: $r=25$, $\xi=0.3$, and $R=100$ for the deterministic stopping set $S=B(0,R)$. }
  \label{fig:PeakAoI_vs_lambda}
\end{figure}

Fig.~\ref{fig:PeakAoI_vs_lambda} depicts the peak AoI as a function of the spatial density for scenarios with and without local observation.
This figure not only illustrates how network densification affects the information freshness, but also highlights the critical role played by the scheduling policy.
Specifically, it shows that the peak AoI always increases with respect to the spatial density, since densifying the network inevitably entails additional interference, thus the SINR is defected and it further hurts the transmission quality across network. Moreover, by employing the locally adaptive scheduling policy at each transmitter, the peak AoI undergoes a substantial discount, and the gain is more pronounced in the dense network scenario.
This is because the interference between neighbors becomes more severe when their mutual distance is reduced, and hence adequately scheduling the channel access patterns of transmitters can prevent the interference from rising too quickly and maintain the peak AoI at a low level.

\section{Conclusion}
In this paper, we proposed a decentralized protocol that allows every transmitter to make transmission decisions based on the observed local information to optimize the information freshness.
Using the concept of stopping sets, we encapsulated the local knowledge from individual nodes in the analytical framework.
The numerical results showed that while the link throughput can be greatly affected by the packet arrival rate, our proposed scheme managed to adapt the transmission to the traffic variation and hence largely reduce the peak AoI.
Moreover, the scheme has also been shown adaptively adjust according to the geographical change of the ambient environment and thus scale well as the network grows in size.

\begin{appendix}
\subsection{Proof of Theorem~\ref{thm:DmSym_PeakAoI}} \label{apx:OptCtrl_PeakAoI}
First of all, we note that under a dominant system, the point process $\Phi$ is stationary.
As such, by substituting \eqref{equ:Dmt_CondThrPut} into the first term of \eqref{prbm:Dom_PeakAoI} and using the mass transportation theorem \cite{BacBla:09}, we obtain the following:
\begin{align}\label{equ:MasTrn}
&\mathbb{E}^0_\Phi \!\! \left[ \frac{ 1 - \xi }{  \gamma^\Phi_0 \mu^\Phi_0 - \xi  } \right] \!=\!  \mathbb{E}^0_\Phi \! \bigg[ \frac{ 1 - \xi }{ \eta_{\mathrm{S}} \! \prod_{j \neq 0} \! \left( 1 \! - \! \frac{ \eta_{\mathrm{S}} }{ 1 + \mathcal{D}_{0j} } \right) \! e^{\frac{ - T r^\alpha }{\rho}} \!\!\!-\! \xi  } \bigg].
\end{align}
Our goal is now to minimize the above expression as a function of $\eta_{\mathrm{S}}$ under the constraint in \eqref{equ:Dom_Rst_SpSt}. To accomplish this target, we separate the denomenator into two sets depending on whether a generic receiver $y_i \in S$ or not. As such, by the strong Markov property, \eqref{equ:MasTrn} can be written as follows
\begin{align}\label{equ:SpltDnm}
&\mathbb{E}^0_\Phi \! \bigg[ \frac{ 1 - \xi }{ \eta_{\mathrm{S}}  \prod_{j \neq 0} \! \left( 1 \! - \! \frac{ \eta_{\mathrm{S}} }{ 1 + \mathcal{D}_{0j} } \right) \! e^{-\frac{ T r^\alpha }{\rho}} \!\!\!-\! \xi  } \bigg]
\nonumber\\
\!=\!& \mathbb{E}^0_\Phi \! \bigg[ \frac{ (1-\xi) e^{ T r^\alpha / \rho } }{ \eta_{\mathrm{S}} \!\!\!\!\!\! \prod\limits_{ \substack{ j \neq 0,  y_j \in S } } \!\!\!\!\!\! ( \, 1 \! - \! \frac{ \eta_{\mathrm{S}} }{ 1 + \mathcal{D}_{0j} } \, ) \, e^{ -\! \int_{ \mathbb{R}^2 \! \setminus \! S } \!\!\! \frac{ \lambda \eta_{\mathrm{S}}  dy }{ 1 + y^\alpha \!/ T r^\alpha } } \!+ \xi e^{ T r^\alpha / \rho } } \bigg].
\end{align}
The minimization of \eqref{prbm:Dom_PeakAoI} now becomes minimizing the above expression with respect to $\eta_{\mathrm{S}}$. Hence, with the help of \eqref{equ:SpltDnm}, we can take the derivative of \eqref{prbm:Dom_PeakAoI} with respect to $\eta_{\mathrm{S}}$ and equate it to zero, which yields the following:
\begin{align} \label{equ:FixPnt_PRF}
\frac{1}{\eta_{\mathrm{S}}} -\!\!\!\!\!\! \sum_{ \substack{ j \neq 0, y_j \in S } }  \frac{1}{ 1 \!+\! \mathcal{D}_{0j}   \!-\! \eta_{\mathrm{S}} } -\!\! \int_{\mathbb{R}^2 \setminus S }\! \frac{ \lambda \,  dy }{ 1 \!+\! y^\alpha \! / T r^\alpha } = 0.
\end{align}
Note that the left hand side (L.H.S.) of the above equation is continuous and decreasing in $\eta_{\mathrm{S}}$ over [0, 1], thus the equation has one solution if \eqref{equ:CndOptl} holds.
Otherwise, if \eqref{equ:CndOptl} does not hold, we have the L.H.S. of \eqref{equ:FixPnt_PRF} being negative which indicates that \eqref{equ:SpltDnm} monotonically increases as a function of $\eta_{\mathrm{S}}$. Hence, the maximum is achieved at $\eta_{\mathrm{S}} = 1$.

\end{appendix}

\bibliographystyle{IEEEtran}
\bibliography{StringDefinitions,IEEEabrv,howard_AoI_Schdl}

\end{document}